\documentclass{scrartcl}
\usepackage{subcaption}
\usepackage{amsmath}
\usepackage{amssymb}

\usepackage{graphicx}

\usepackage{hyperref}
\usepackage{doi}
\usepackage[numbers]{natbib}
\usepackage{float}

\usepackage{color} 

\usepackage{setspace} 
\usepackage{pgf,tikz}
\usetikzlibrary{external,plotmarks,spy}
\begin{document}

\begin{flushleft}
{\Large
\textbf{Modeling the effects of variable feeding patterns of larval ticks on the transmission of \textit{Borrelia lusitaniae} and \textit{Borrelia afzelii}}
}
% Insert Author names, affiliations and corresponding author email.
\\
Luca Ferreri$^{1,\ast}$, Silvia Perazzo$^{2}$, Ezio Venturino$^{2}$, Mario Giacobini$^{1,3}$, Luigi Bertolotti$^{1}$, Alessandro Mannelli$^{1}$\\
\bfseries{1} Department of Veterinary Sciences, University of Torino, Italy
\\
\bfseries{2} Department of Mathematics, University of Torino, Italy
\\
\bfseries{3} Molecular Biotechnology Centre, University of Torino, Italy
\\
$\ast$ E-mail: \href{mailto:luca.ferreri@unito.it}{luca.ferreri@unito.it}
\end{flushleft}

\begin{abstract}
Spirochetes belonging to the \textit{Borrelia burgdoferi} sensu lato (sl) group cause Lyme Borreliosis (LB), which is the most commonly reported vector-borne zoonosis in Europe. \textit{B. burgdorferi} sl is maintained in nature in a complex cycle involving \textit{Ixodes ricinus} ticks and several species of vertebrate hosts. The transmission dynamics of \textit{B. burgdorferi} sl is complicated by the varying competence of animals for different genospecies of spirochetes that, in turn, vary in their capability of causing disease. In this study, a set of difference equations simplifying the complex interaction between vectors and their hosts (competent and not for \textit{Borrelia}) is built to gain insights into conditions underlying the dominance of \textit{B. lusitaniae} (transmitted by lizards to susceptible ticks) and the maintenance of \textit{B. afzelii} (transmitted by wild rodents) observed in a study area in Tuscany, Italy. Findings, in agreement with field observations, highlight the existence of a threshold for the fraction of larvae feeding on rodents below which the persistence of \textit{B. afzelii} is not possible. Furthermore, thresholds change as nonlinear functions of the expected number of nymph bites on mice, and the transmission and recovery probabilities. In conclusion, our model provided an insight into mechanisms underlying the relative frequency of different \textit{Borrelia} genospecies, as observed in field studies.
\end{abstract}

	\section{Introduction}
		Lyme borreliosis (LB), caused by spirochetes belonging to the \textit{Borrelia burgdorferi sensu lato} (sl) group, is the most commonly reported vector-borne zoonosis in temperate climates. In Europe, \textit{B. burgdorferi} sl is maintained in transmission cycles involving the tick \textit{Ixodes ricinus} (in combination with \textit{I. persulcatus} in areas in North Eastern Europe), and several species of vertebrate reservoir hosts that can be infected by ticks and that, in turn, are able to transmit the infection to other susceptible ticks \citep{gernetal1998,gernhumair1998}. A distinct feature of the transmission cycle in Europe is a certain degree of reservoir host-specificity of different genospecies of \textit{B. burgdorferi} sl. In fact, among pathogenic genospecies, rodents and other small mammals transmit \textit{B. burgdorferi} sensu stricto and \textit{B. afzelii}, birds are reservoirs for \textit{B. garinii}, whereas lizards transmit \textit{B. lusitaniae} (detected in human patients especially in Portugal \citep{Collares-Pereira}).
		
		Although \textit{B. burgdorferi} sl is widespread through the geographic range of \textit{I. ricinus}, the prevalence of infection in host-seeking ticks, and the relative frequency of different genospecies may vary within short distances. The composition of populations of vertebrate hosts, which are characterized by varying reservoir competence for \textit{B. burgdorferi} sl genospecies, might play a major role in the ecological processes underlying such a variability. More specifically, geographic variations of the intensity of transmission of \textit{B. burgdorferi} sl genospecies might be the result of the relative contribution, by each vertebrate host species, to the overall infection of susceptible larvae; this depends upon the host's population density, the average number of larvae per individual of that species, and the host's infectivity to larvae (the fraction of larvae that acquire the infection after feeding on vertebrates of a certain species) \citep{mather1989}.
		
		Large mammals, such as deer and other wild and domestic ungulates, are considered unable to serve as reservoirs for \textit{B. burgdorferi} sl. On the other hand, they play a major role as hosts for adult ticks. Consequently, the effects of population densities of non competent hosts on \textit{B. burgdorferi} sl transmission dynamics  cannot be easily predicted, due to their potential, contrasting effects of dilution and amplification of transmission \citep{Keesing}. More specifically, non-competent hosts may reduce nymphs infection prevalence by feeding relatively large proportions of larvae that moult to non-infected nymphs, and by diverting ticks from competent reservoirs, resulting in a dilution effect. Nevertheless, non-competent hosts may feed large numbers of ticks and, therefore, augment the vector population. This may result in a larger probability that ticks feed on an infected host with the consequent amplification of \textit{B. burgdorferi} sl transmission.
		
		In the European situation, an animal species may serve as a reservoir host and, therefore, amplify the transmission of certain \textit{B. burgdorferi} sl genospecies. It may simultaneously affect transmission of other genospecies through dilution and/or vector augmentation (see \citep{Mannelli} for a summary of transmission in multi-host systems). These complex mechanisms might, at least in part, explain why several genospecies may thrive in certain areas, whereas, at other locations, one genospecies might be dominant while the others are rare.
		
		In a study area on Le Cerbaie Hills, in Tuscany (Central Italy), previous studies hypothesized that lizards were responsible for the maintenance of \textit{B. lusitaniae} as the dominant genospecies and, at the same time, reduced the transmission of other genospecies through dilution. A simple mathematical model indicated that, on Le Cerbaie, persistence of \textit{B. afzelii} ($R_0 > 1$) was only possible under conditions of relatively large density of mice reservoir hosts (\textit{Apodemus} spp) and large attachment rate of \textit{I. ricinus} nymphs to mice \citep{ragaglietal2011}. Indeed, mouse population fluctuations, and  the frequency of bites by immature \textit{I. ricinus} were recognized as key factors affecting these hosts' specificity and, consequently, the intensity of transmission of \textit{B. afzelii} \citep{Mannelli}. Our work fits within this context adding some details (the dynamical model and the explicit definition of densities of vertebrate hosts) in order to increase the accuracy of the model and to further confirm the previous finding.
		
		In this study, we build a simple dynamical model to study the transmission of \textit{B. burgdorferi} sl genospecies under variable scenarios regarding the relative contribution of different hosts to the feeding of larval ticks, and the frequency of bites by infectious nymphs on the same vertebrates. After the model's general formulation, we use it to gain an insight into mechanisms underlying the observed variations in the prevalence of \textit{B. burgdorferi} sl genospecies in ticks and hosts on Le Cerbaie.  Specifically, we explore conditions leading to persistence or extinction of \textit{B. lusitaniae} and \textit{B. afzelii} in the study area.

	\section{Methods}
		We used a set of recurrence relations to describe the complex interactions underlying the transmission of  different genospecies of \textit{B. burgdoferi} sl among \textit{I. ricinus} immature stages (larvae and nymphs) and vertebrate hosts. Adult ticks were not included in our model. As a consequence, the vector augmentation effect of animals serving as hosts for this stage was not considered. Therefore, the focus of our study was limited to the contribution of each host species to larval feeding, and to the frequency of bites by infectious nymphs on different hosts and their effects on the persistence of \textit{B. burgdoferi} sl genospecies. Furthermore, we only considered the infection of larvae through feeding on systemically infected hosts -- the most common transmission route of \textit{B. burgdorferi} sl in nature \citep{Mannelli,gern1994}. Accordingly, we disregarded transovarial infection of larvae from female ticks of the previous generation, and transmission of spirochetes among ticks feeding in close proximity on the host's skin (transmission via co-feeding), \citep{matuschkaetal1998,patrican1997,Ogden1997,randolphrogers2006}.
		
		Feeding larvae may acquire a certain genospecies of \textit{B. burgdorferi} sl depending on the specific reservoir competence of the parasitized host (\textit{B. lusitaniae} can be acquired by feeding on lizards, \textit{B. afzelii} by feeding on mice). On the other hand, larvae feeding on non-competent hosts (such as deer), do not acquire these agents. Infected, fed larvae molt into infected nymphs, which are subsequently able to transmit the infection to susceptible hosts.
		
		Across most of the geographic range of \textit{I. ricinus}, including Le Cerbaie, nymphs are active before larvae during the same year \citep{Bisanzio2008}. Such a seasonal pattern is particularly favorable to the maintenance of \textit{B. burgdorferi} sl. In fact, in Spring, spirochetes are transmitted by nymphs to susceptible hosts that, in turn, develop a systemic infection and are able to transmit it to susceptible larvae in the following Summer \citep{Mannelli}. Therefore, in our model, we consider a time-step, $\Delta t$, of six months, and we assume that, for each year, larvae  feed on hosts only during the second semester (July-December), whereas nymphs feed during the first semester only (January-June). Under such conditions, competent hosts ensure transmission of the infection between nymphs and larvae belonging to different tick generations, allowing the maintenance of \textit{B. burgdorferi}.
		\subsection{Recurrence relations}		
			In order to understand the role of varying host populations on the endemic condition of the genospecies we introduce a parameter $h_{S}$, the specificity of host-species $S$ on feeding larvae, denoting the fraction of larvae feeding on a particular host species, $S$. In fact, we have $h_L$, for those feeding on lizards, $h_R$, for those feeding on rodents, and $h_H$, for those feeding on other, non-competent hosts (thus $h_L+h_R+h_H=1$). Furthermore, we assume that the vector and host populations are constant in time.
		
			Now, since we do not assume any correlation between the probability for a nymph to feed on a rodent and the probability that the nymph was already infectious before the moult, the prevalence of \textit{B. afzelii}, $\pi_a$, among nymphs feeding on rodents at the beginning of time $t+\Delta t$ is equal to the prevalence among larvae that have completed a blood meal at the end of time $t$. In particular, assuming frequency-dependent transmisson \citep{begon2002}, the latter is a function of the prevalence of \textit{B. afzelii}, $p_a(t)$, among rodents on which larvae fed and is given by the probability that a feeding larvae gets infected at the end of its blood meal. That is
			\begin{equation}
				\pi_a(t+\Delta t)=\beta_{RT}\cdot\delta_T(t)\cdot h_R\cdot p_a(t)
			\end{equation}
			where $\delta_T$ is the Kronecker delta, which is one if $t$ is the semester of activity of larvae (i.e. second half of the year) and zero otherwise (i.e. January-June), and $\beta_{RT}$ is the probability that a larva biting an infectious rodent becomes infectious. In a similar way we depict the prevalence of \textit{B. lusitaniae} among feeding nymphs, $\pi_l$:
			\begin{equation}
				\pi_l(t+\Delta t)=\beta_{LT}\cdot\delta_T(t)\cdot h_L\cdot p_l(t)
			\end{equation}
			where $\beta_{LT}$ is the probability that a larva biting an infectious lizard gets the infection and $p_l(t)$ is the prevalence of \textit{B. lusitaniae} among lizards. On the other hand, the prevalence of \textit{B.afzelii}, $p_a(t+\Delta t)$, among rodents is nothing but a function of those mice still infected from previous time-step, i.e. $(1-\mu_R)\cdot p_a(t)$, where $\mu_R$ is the probability for a mouse to die and to be replaced (in order to keep the population stable in time), and those rodents that get infected as consequences of infecting bites at the previous time-step. We do not consider the possibility of hosts recovery due to the low probability of this event, \citep{gern1994,kurtenbach2002,kurtenbach1994,zhong1997}. Moreover, defining $\beta_{NR}$ as the probability that an infected nymph bite is infective and $K_R$ as the expected number of nymph bites for a mouse in a time step $\Delta t$, we could expect a number of $\beta_{NR}\cdot K_R\cdot \pi_a(t)$ potentially infectious bites for mouse. Therefore, since one potentially infectious bite comes independently from the others, we could assume the random variable ``\textit{number of infectious nymph bites for mouse at time step $\left[t,t+\Delta t\right)$}'' is Poisson distributed with mean $\beta_{NR}\cdot K_R\cdot \pi_a(t)$. As consequence, the probability for a susceptible mouse to be infected by at least one nymph is nothing but $1-e^{-\beta_{NR}\cdot K_R\cdot \pi_a(t)}$. Summarizing
			\begin{equation}
				p_a(t+\Delta t)=\left(1-\mu_R\right)\cdot p_a(t)+\left[1-p_a(t)\right]\cdot\left[1-e^{-\beta_{NR}\cdot K_R\cdot \pi_a(t)}\right].
			\end{equation}
			Similarly, the prevalence of \textit{B. lusitaniae}, $p_l$, among lizards is
			\begin{equation}p_l(t+\Delta t)=\left(1-\mu_L\right)\cdot p_l(t)+\left[1-p_l(t)\right]\cdot\left[1-e^{-\beta_{NL}\cdot K_L\cdot \pi_l(t)}\right],\end{equation}
			where $\mu_L$ is the probability that a lizard dies and is replaced (to have stable lizard population), $\beta_{NL}$ is the probability that an infectious nymph bite is infective for a lizard and $\beta_{NL}\cdot K_L\cdot \pi_l(t)$ is the expected number of potentially infectious bites for lizards in $\left[t,t+\Delta t\right)$.

			It is important  to underline that the two equations describing the prevalence of \textit{B. lusitaniae} and those describing the prevalence of \textit{B.afzelii} are coupled together only through the constraint $h_L+h_R+h_H=1$.
		\subsection{Equilibria}
			We investigate the equilibria of the prevalence of both genospecies after two time-steps, i.e. one year. That means that for genospecies $g$ we impose $p_g(t+2)=p_g(t)=p_g$. Hence, equilibria of \textit{B. afzelii} among rodents is solution of the following equation:
			\begin{equation}
					p_a=f(p_a)=\left(1-\mu_R\right)^2\cdot p_a+\left[1-\left(1-\mu_R\right)p_a\right]\cdot\left\{1-e^{-\beta_{NR}\cdot\beta_{RT}\cdot K_R\cdot h_R\cdot p_a}\right\}\label{eq_eq_a}
			\end{equation}
			which, at least to our knowledge, can not be solved analytically. However, we have that in the parameters' domain:
			\begin{itemize}
				\item $f(x)$ is a continuous function for $x\in[0,1]$, 
				\item $0$ is a solution, 
				\item $f''(x)<0$,
				\item $f(1)=(1-\mu_R)^2+\mu_R\left\{1-e^{-\beta_{NR}\cdot\beta_{RT}\cdot K_R\cdot h_R}\right\}<1-\mu_R+\mu_R^2$, which for $\mu_R \in (0,1)$ is less than $1$. 
			\end{itemize}
			Therefore, if $f'(0)>1$ there exists only one non-zero solution $\hat{x}\in(0,1)$. Otherwise, only the \textit{B. afzelii}-free equilibrium could exist. After some algebraic manipulation $f'(0)>1$ can be written as:
			\begin{equation}
				h_R>\frac{1-\left(1-\mu_R\right)^2}{\beta_{NR}\cdot\beta_{RT}\cdot K_R}=T_R.\label{th_eq_a}
			\end{equation}
			Similarly, the condition for the persistence of \textit{B. lusitaniae} is
			\begin{equation}
				h_L>\frac{1-\left(1-\mu_L\right)^2}{\beta_{NL}\cdot\beta_{LT}\cdot K_L}=T_L.\label{th_eq_l}
			\end{equation}
			\begin{figure}
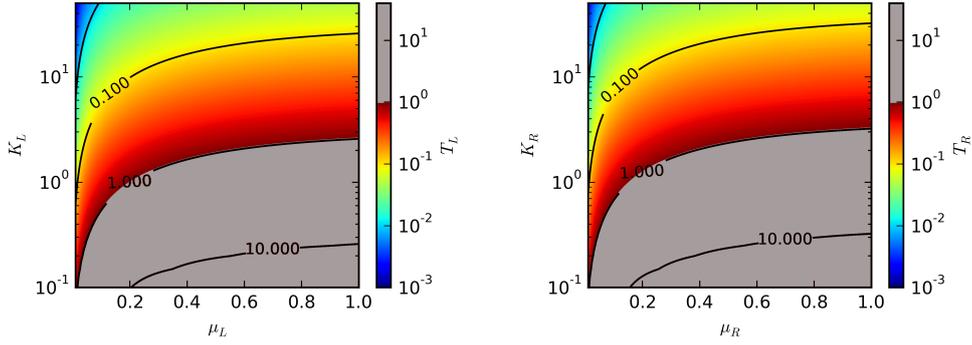

				\centering
				\begin{subfigure}[b]{.45\textwidth}
					\includegraphics[width=\textwidth]{picture/T_L_new_v2.pdf}
				\end{subfigure}
				\begin{subfigure}[b]{.45\textwidth}
					\includegraphics[width=\textwidth]{picture/T_R_new_v2.pdf}
				\end{subfigure}
				\caption{Left: values of $T_L$ as function of $\mu_L$ and $K_L$ while $\beta_{NL}=0.67$ and $\beta_{LT}=0.5$. Right: values of $T_R$ as function of $\mu_R$ and $K_R$ while $\beta_{NR}=0.67$ and $\beta_{RT}=0.4$. Gray values are for unfeasible regions, i.e. $T_L\geq 1$ and $T_R\geq 1$}\label{hlhr}
			\end{figure}
		We explore the thresholds $T_R$ and $T_L$ in Fig. \ref{hlhr}. Now, we could be interested to have the prevalence at equilibria as a function of the fraction of larvae feeding on a specific host. As written before, an analytical function, at least to our knowledge, can not be achieved. However, in \ref{function}, we give two equivalent methods to obtain such a function numerically and consequently to investigate it. 
	\section{Results}
		\subsection{Equilibria}
\begin{figure}[h!]
			\centering{
			\begin{tikzpicture}[x=5cm,y=5cm]
				\fill[color=red!30] (0,0)--(1,0)--(0,1);
				\draw[->](-0.05,0)--(1.25,0)node[anchor=north]{$h_R$};
				\draw[->](0,-0.05)--(0,1.25)node[anchor=east]{$h_L$};
				\draw(-0.05,1)node[anchor=east]{$1$}--(0,1);
				\draw(1,-0.05)node[anchor=north]{$1$}--(1,0);
%				\draw[dashed] (0.25,-0.05)node[anchor=north]{$\frac{1-\left(1-\mu_R\right)^2}{\beta_{NR}\cdot\beta_{RT}\cdot K_R}$}--(0.25,1.25);
%				\draw[dashed] (-0.05,.35)node[anchor=east]{$\frac{1-\left(1-\mu_L\right)^2}{\beta_{NL}\cdot\beta_{LT}\cdot K_L}$}--(1.25,0.35);
				\draw[dashed] (0.25,-0.05)node[anchor=north]{$T_R$}--(0.25,1.25);
				\draw[] (0.65,-0.025)node[anchor=north]{$\left(1-h_H\right)$}--(0.65,0);
				\draw[] (-0.025,.65)node[anchor=east]{$\left(1-h_H\right)$}--(0,0.65);
				\draw[dashed] (-0.05,.35)node[anchor=east]{$T_L$}--(1.25,0.35);
				\draw[color=blue,dotted,line width=1] (.65,0)--(0,.65);
				\draw (1,0)--(0,1);
%				\node[fill=red!30] at (0.6,0.8){};
%				\node[anchor=west]at (0.65,0.8){$h_R+h_L+h_H=1$};
				\node[label=above:$E_0$] at (0.1,0.1){};
				\node[label=above:$E_a$] at (0.6,0.1){};
				\node[label=above:$E_l$] at (0.1,0.6){};
				\node[label=above:$E_{a+l}$] at (0.375,0.35){};
			\end{tikzpicture}}\caption{Different equilibria scenarios as function of the parameters considered. The thresholds $T_R$ and $T_L$ were defined in Eq.~(\ref{th_eq_a}) and (\ref{th_eq_l}) respectively. The highlighted area is the parameters' feasible space. Dotted line shows the range of values that the couple $(h_R,h_L)$ could assume when we fix the value of $h_H$.}\label{schema}
			\end{figure}
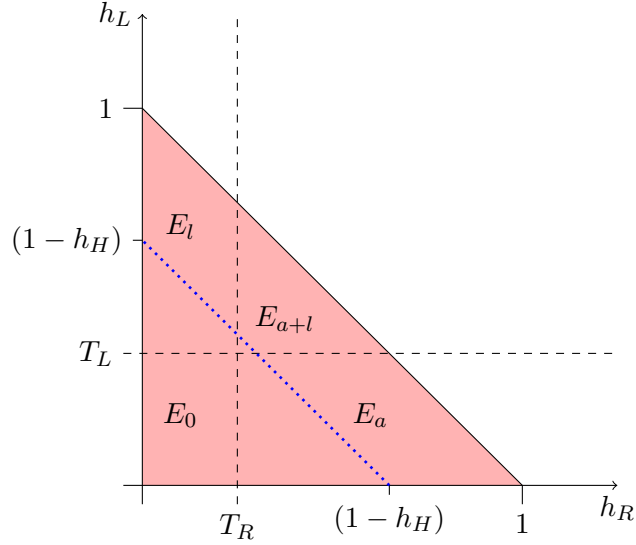

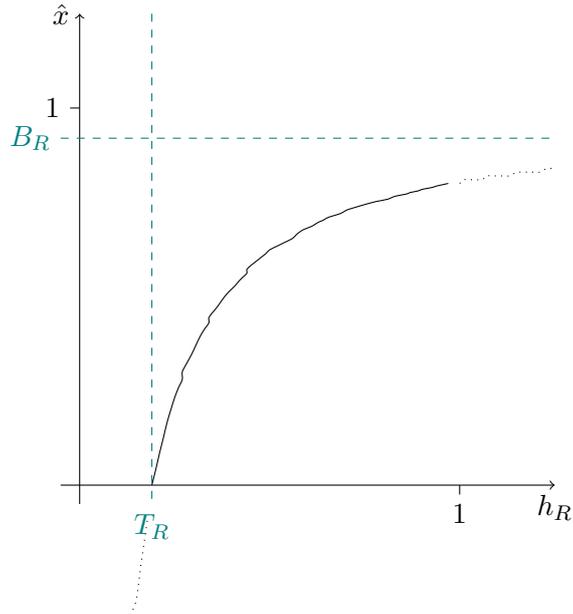
\begin{figure}[h!]
				\centering{
				\begin{tikzpicture}[x=5cm,y=5cm]
				\draw[->](-0.05,0)--(1.25,0)node[anchor=north]{$h_R$};
	%			\node[anchor=south,color=magenta] at(1.25,0){$h_L$};
				\draw[->](0,-0.05)--(0,1.25)node[anchor=east]{$\hat{x}$};
	%			\node[anchor=west,color=magenta] at(0,1.25){$\hat{y}$};
	%			\draw[domain=0.11:1]plot(\x,{-(\x+.9)^(-9)+0.9});
%				\draw[dotted] plot[smooth]coordinates{(0.14,-0.33) (0.14,-0.32) (0.14,-0.32) (0.15,-0.3) (0.15,-0.25) (0.15,-0.24) (0.16,-0.22) (0.16,-0.21) (0.16,-0.2) (0.16,-0.18) (0.16,-0.17) (0.17,-0.15) (0.17,-0.14) (0.17,-0.11) (0.18,-0.08) (0.18,-0.07) (0.18,-0.06) (0.18,-0.06) (0.18,-0.05) (0.18,-0.04)};
				\draw[dotted] plot[smooth]coordinates{(0.14,-0.33) (0.15,-0.3)(0.16,-0.22) (0.17,-0.15) (0.18,-0.08) };
%				\draw plot[smooth,tension=0.1]coordinates{ (0.19,0) (0.2,0.03) (0.2,0.04) (0.2,0.06) (0.2,0.06) (0.2,0.07) (0.21,0.09) (0.22,0.13) (0.22,0.14) (0.22,0.16) (0.23,0.17) (0.23,0.17) (0.23,0.18) (0.23,0.18) (0.23,0.19) (0.24,0.2) (0.24,0.21) (0.25,0.24) (0.25,0.24) (0.25,0.25) (0.25,0.25) (0.26,0.26) (0.26,0.26) (0.26,0.27) (0.27,0.29) (0.27,0.3) (0.28,0.31) (0.28,0.32) (0.28,0.33) (0.29,0.34) (0.29,0.35) (0.3,0.36) (0.3,0.36) (0.3,0.37) (0.3,0.37) (0.31,0.38) (0.31,0.39) (0.31,0.39) (0.32,0.4) (0.32,0.41) (0.33,0.42) (0.33,0.43) (0.34,0.44) (0.34,0.44) (0.34,0.45) (0.35,0.45) (0.35,0.46) (0.35,0.46) (0.36,0.47) (0.36,0.47) (0.37,0.49) (0.38,0.5) (0.39,0.51) (0.39,0.51) (0.39,0.52) (0.39,0.52) (0.4,0.52) (0.4,0.53) (0.4,0.53) (0.41,0.53) (0.41,0.54) (0.42,0.54) (0.42,0.55) (0.43,0.55) (0.43,0.56) (0.44,0.57) (0.44,0.57) (0.45,0.58) (0.46,0.59) (0.47,0.6) (0.48,0.61) (0.49,0.61) (0.49,0.61) (0.5,0.62) (0.5,0.62) (0.5,0.62) (0.52,0.63) (0.52,0.64) (0.54,0.65) (0.56,0.66) (0.57,0.66) (0.58,0.67) (0.59,0.68) (0.61,0.69) (0.61,0.69) (0.62,0.69) (0.63,0.7) (0.64,0.7) (0.64,0.7) (0.65,0.71) (0.66,0.71) (0.66,0.71) (0.66,0.71) (0.68,0.72) (0.69,0.72) (0.7,0.73) (0.7,0.73) (0.71,0.73) (0.71,0.73) (0.73,0.74) (0.75,0.74) (0.78,0.75) (0.81,0.76) (0.83,0.77) (0.86,0.77) (0.87,0.78) (0.89,0.78) (0.9,0.78) (0.9,0.79) (0.91,0.79) (0.91,0.79) (0.93,0.79) (0.93,0.79) (0.94,0.79) (0.97,0.8)};
				\draw plot[smooth]coordinates{ (0.19,0) (0.2,0.04) (0.21,0.084) (0.22,0.124) (0.23,0.166) (0.24,0.2) (0.25,0.232) (0.26,0.258) (0.27,0.28) (0.27,0.3) (0.28,0.32) (0.29,0.338) (0.3,0.358) (0.31,0.38) (0.32,0.4) (0.33,0.416) (0.34,0.43) (0.34,0.444) (0.35,0.458) (0.36,0.47) (0.37,0.484) (0.38,0.498) (0.39,0.51) (0.4,0.52) (0.41,0.53) (0.42,0.542) (0.43,0.552) (0.44,0.562) (0.44,0.572) (0.45,0.582) (0.46,0.59) (0.47,0.598) (0.48,0.606) (0.49,0.614) (0.5,0.624) (0.52,0.634) (0.54,0.644) (0.56,0.654) (0.57,0.664) (0.58,0.672) (0.59,0.678) (0.61,0.686) (0.62,0.692) (0.63,0.698) (0.64,0.702) (0.65,0.708) (0.66,0.712) (0.68,0.718) (0.69,0.722) (0.7,0.728) (0.71,0.732) (0.73,0.738) (0.75,0.744) (0.78,0.752) (0.81,0.758) (0.83,0.766) (0.86,0.772) (0.87,0.776) (0.89,0.78) (0.9,0.784) (0.91,0.786) (0.93,0.79) (0.94,0.793333) (0.97,0.8)};
%				 \draw[dotted] plot[smooth]coordinates{(1,0.8) (1.01,0.81) (1.02,0.81) (1.04,0.81) (1.04,0.81) (1.06,0.81) (1.07,0.82) (1.09,0.82) (1.11,0.82) (1.13,0.82) (1.15,0.83) (1.16,0.83) (1.17,0.83) (1.18,0.83) (1.19,0.83) (1.19,0.83) (1.2,0.83) (1.21,0.83) (1.23,0.84) (1.24,0.84) (1.25,0.84)};
				 \draw[dotted] plot[smooth]coordinates{(1,0.8) (1.01,0.81) (1.02,0.81) (1.04,0.81) (1.06,0.81) (1.07,0.82) (1.09,0.82) (1.11,0.82) (1.13,0.82) (1.15,0.83) (1.16,0.83) (1.17,0.83) (1.18,0.83) (1.19,0.83) (1.2,0.83) (1.21,0.83) (1.23,0.84) (1.24,0.84) (1.25,0.84)};
	%			\draw[domain=0.075:1.25,dotted]plot(\x,{-(\x+0.9)^(-9)+0.9});
	%			\draw[domain=0.44:1,color=magenta]plot(\x,{-(\x+.6)^(-13)+0.6});
	%			\draw[domain=0.4:1.25,dotted,color=magenta]plot(\x,{-(\x+0.6)^(-13)+0.6});
				\draw (1,0)--(1,-0.025)node[anchor=north]{$1$};
				\draw (0,1)--(-0.025,1)node[anchor=east]{$1$};
				\draw[color=teal,dashed] (0.19,1.25)--(0.19,-0.05)node[anchor=north]{$T_R$};
				\draw[color=teal,dashed] (-0.05,0.92)node[anchor=east]{$B_R$}--(1.25,0.92);
	%			\draw[color=magenta] (0.44,-0.025)--(0.44,1.25)node[anchor=south]{$-\frac{1-\left(1-\mu_R\right)^2}{\beta_{RT}\cdot K_R\ln(1-\beta_{NR})}$};
	%			\draw[color=magenta] (-0.05,0.6)node[anchor=east]{$-\frac{(1-\mu_R)}{\mu_R^2-\mu_R-1}$}--(1.25,0.6	);
				\end{tikzpicture}}
				\caption{\textit{B. afzelii} prevalences among rodents at equilibria, $\hat{x}$, as function of the fraction of feeding-ticks feeding on rodents, $h_R$ (solid line in the feasibility region). The threshold $T_R$ was defined in Eq.~(\ref{th_eq_a}), while the upper bound of $\hat(x)$ is given by $B_R=\frac{1}{1+\mu_R-\mu_R^2}$.}\label{prev}
				\end{figure}

			According to the parameters considered, our system has different steady states. We schematize them on Fig. \ref{schema}. In particular, possibilities are:
			\begin{itemize}
				\item $E_0$ the pathogens-free equilibrium, reached when both $h_R<T_R$ and $h_L<T_L$ or in other words when fraction of larvae that feed on competent hosts is too low to maintain the pathogen,
				\item $E_a$ the \textit{B. afzelii}-only equilibrium reached when $h_R>T_R$ but $h_L<T_L$, i.e. fraction of larvae that feed on rodents is enough to reach the endemic equilibrium of \textit{B. afzelii}-only,
				\item $E_l$ the \textit{B. lusitaniae}-only equilibrium reached when $h_R<T_R$ but $h_L>T_L$, i.e. fraction larvae feed on lizards is enough to reach the endemic equilibrium of \textit{B. lusitaniae}-only,
				\item $E_{a+l}$  the both genospecies equilibrium reached when $h_R>T_R$ and $h_L>T_L$, i.e. fraction larvae feed on rodents and lizards is enough to reach the endemic equilibrium of both genospecies.
			\end{itemize}
		Moreover, \textit{B. afzelii} prevalence among rodents at equilibria, $\hat{x}$, as function of the fraction of feeding-ticks feeding on rodents is sketched in Fig. \ref{prev}. In a similar manner we depict \textit{B. lusitaniae} prevalence among lizards. For \textit{B. afzelii} when the value of $h_R$ is above the threshold $T_R$ the prevalence of the genospecies among rodent at equilibria grows super-linearly as the $h_R$ increases. In addition, we have that the prevalence is bounded by $B_R=\frac{1}{1+\mu_R-\mu_R^2}$.
		\subsection{Equilibrium analysis}
			\textit{B. afzelii} prevalence among rodents at equilibrium is \textit{asymptotically stable}, \citep{castillo}, when it exists. Indeed, since $f(\hat{x})=\hat{x}$ we have that $f'(\hat{x})<1$. Furthermore,
			\begin{multline*}f'(x)=\left(1-\mu_R\right)^2-\left(1-\mu_R\right)\cdot\left(1-e^{-\beta_{NR}\cdot\beta_{RT}\cdot K_R\cdot h_R\cdot x}\right)\\+\left[1-\left(1-\mu_R\right)\cdot x\right]\cdot \beta_{NR}\cdot\beta_{RT}\cdot K_R\cdot h_R\cdot e^{-\beta_{NR}\cdot\beta_{RT}\cdot K_R\cdot h_R\cdot x}>\\-\left(1-\mu_R\right)\cdot\left(1-e^{-\beta_{NR}\cdot\beta_{RT}\cdot K_R\cdot h_R\cdot x}\right)>-1\end{multline*}
			Within these conditions, the other equilibrium, $p_a=0$ i.e. the \textit{B. afzelii}-free equilibria, is  not stable. On the other hand, when the condition reported in eq.\ref{th_eq_a} for the existence of non-zero \textit{B. afzelii} prevalences among rodent does not hold, the \textit{B. afzelii}-free equilibria is \textit{asymptotically stable}. The same behaviors are also observed for the endemicity and stability of \textit{B. lusitaniae}.
		\subsection{Application of the model to Le Cerbaie Hills' scenario}
			To apply the model to the ecological scenarios which were previously observed in field studies on Le Cerbaie Hills, we adopted the following parameters:
			\begin{itemize}
				\item probability of transmission of \textit{B. afzelii} from rodent to \textit{I. ricinus} larvae, $\beta_{RT} = 0.4$, \citep{Hanincova2003,Randolph1995};
				\item probability of transmission of \textit{B. lusitaniae} from lizards to \textit{I. ricinus} larvae, $\beta_{LT} = 0.5$, \citep{ragaglietal2011};
				\item probability of transmission of \textit{B. burgdorferi} sl from \textit{I. ricinus} nymphs to hosts (lizards or rodents), $\beta_{NL} = \beta_{NR} = 0.67$, \citep{ragaglietal2011};
				\item assuming an exponential distribution for the waiting time of the death event of an agent, we can estimate the removal probability, $\mu$, from the average lifespan, $\ell$, as follows
\[1-\exp\left(-\left(\frac{1}{\ell}\right)\right).\] 
Hence, assuming the rodent life span to be one year, thus $\ell=2$, we have that $\mu_R=1-\exp(-.5)$\citep{topi1,topi2};
				\item accordingly, assuming lizard life span to be five years \citep{lucertole}, thus $\ell=10$, we have that $\mu_L=1-\exp(-.1)$;
				\item mean number of bites by \textit{I. ricinus} nymphs per host per year was calculated from the mean number of nymphs which were found attached on hosts of a given species examined in the field, after assuming that nymphs were active during two months ($60$ days) in a year, and that each nymph remained attached to the host for four days. Therefore, 
				\[K_H=\left(\textrm{mean number of nymphs on host of species } H\right)\times 60/4.\] 
				Based upon field observations on Le Cerbaie \citep{ragaglietal2011}, two scenarios were considered for this parameter:
				\begin{itemize}
					\item in scenario I, $K_R = 3.825$ and $K_L = 23.52$;
					\item in scenario II, $K_R = 0.27$ and $K_L =48.87$; 
				\end{itemize}
				Length of two months for nymphs activity is arbitrarily assumed based on the field observations. Furthermore, we observe that our results are robust for the nymphs activity span between 25 and 180 days, Fig.\ref{hlhr}.
			\end{itemize}
			By using the above parameters we found that, in scenario I, persistence of \textit{B. afzelii} at the endemic status is only possible when the fraction of larvae that feed on rodents, $h_R$, is  at least $62\%$, whereas stable persistence of \textit{B. lusitaniae} only needs  $h_L$ to be larger than $10\%$. On the other hand, our results show that in scenario II,  \textit{B. afzelii} does not reach the endemic status under any value of $h_R$. Conversely, in scenario II, $h_L > 5\%$ would be sufficient to allow \textit{B. lusitaniae} to persist.
		
	\section{Discussion and Conclusions}
		In this study we defined a dynamical model of the transmission process of \textit{B. lusitaniae} and \textit{B. afzelii}. This is an improvement of the model by \citet{ragaglietal2011}, where the model did not explicitly take into account the densities of vertebrate hosts. Indeed, in the present model we more accurately described the transmission dynamics in their temporal aspects. The dynamical model presented in this study was specifically geared to analyse the effects of feeding patterns of immature \textit{I. ricinus} on vertebrate hosts serving as reservoirs for different \textit{B. burgdorferi} sl genospecies, and to better understand mechanisms underlying variations of the intensity of transmission of these agents. Specifically, the dominance of \textit{B. lusitaniae}, and the variable persistence of \textit{B. afzelii} on Le Cerbaie Hills were analysed using infestation patterns of lizards and small rodents.
		Lizards are important components of wildlife in the warm and relatively dry sub Mediterranean habitat of Le Cerbaie, where \textit{I. ricinus} finds enough moisture for its development and activity. Since they favor both vertebrate reservoir hosts as well as arthropod vectors, these environmental conditions are particularly suitable for the establishment of  \textit{B. lusitaniae} foci. Indeed, we identified lizards' ability to serve as hosts for the tick's immature stages, together with their relatively long lifespan, as key factors enhancing \textit{B. lusitaniae} transmission dynamics.
		
		On the contrary, we have shown that \textit{B. afzelii} can only persist on Le Cerbaie when $h_R$, the fraction of \textit{I. ricinus} larvae feeding on \textit{Apodemus} spp (the reservoir hosts for this genospecies), and the mean number of bites by \textit{I. ricinus} nymphs per mouse per year, $K_R$, are at their maximum observed values. $h_R$ is the result of mouse population density and of the number of \textit{I. ricinus} larvae per mouse, relatively to the same parameters for all hosts on which this tick stage may feed. Mouse population fluctuations may, therefore, directly influence $h_R$ and, consequently, our model results. Accordingly, in the field, \textit{B. afzelii} was not found under conditions of low \textit{Apodemus} spp density, and it was only found in ticks from \textit{Apodemus} spp when the rodents' population reached relatively high levels \citep{ragaglietal2011}.

		It is important to emphasize out that parameter $K_R$ ($K_L$) is probably associated with $h_R$ ($h_L$). However, our model does not make any assumption in this direction and thus it is open to any real scenario. Extensive research on the relation between $K_R$ ($K_L$) and $h_R$ ($h_L$) will be the subject of future works.
		
		The role of $K_R$ on \textit{B. afzelii} transmission is supported by previous research. In fact, in the UK, the rarity of this genospecies was attributed to the low attachment rate of \textit{I. ricinus} nymph on \textit{Apodemus} spp \citep{Gray1999}. Variations of $K_R$ on Le Cerbaie (which we considered in different modeling scenarios) could, hypothetically, be explained by interactions between host-seeking behavior of \textit{I. ricinus} nymphs, and the circadian activity of small rodents. Specifically, in dry upland habitat, where \textit{B. afzelii} was found in 2006, nymphs might be more active during the night, when there is more humidity and \textit{Apodemus} spp are also active, leading to an enhanced probability of encounter between nymphs and mice \citep{Mannelli}. 
		
		It was previously suggested that lizards might be responsible for the dilution of the transmission of \textit{B. afzelii} on Le Cerbaie, by diverting immature ticks from the \textit{Apodemus} spp mice - the competent reservoir hosts for this genospecies \citep{amoreetal2007}. Such an hypothesis would be in agreement with results of the present model, since such a diversion would lead to reduced  $h_R$ and $K_R$, which we have shown as critical in the persistence or extinction of \textit{B. afzelii}. Field studies in the Mediterranean Region have shown that \textit{B. lusitaniae} was the dominating genospecies in \textit{I. ricinus} \citep{DeMichelis2000,Younsi2001,Sarih2003}, and this might be due to the dominance of lizards as vertebrate hosts for ticks at the same locations. Furthermore, even in Germany, lizards are considered as able to negatively affect the transmission of genospecies other than \textit{B. lusitaniae} \citep{Richter2013}.
		
		The dilution role of lizards cannot, however, be demonstrated based on our study. Recent research, in fact, suggested that even animal species serving as hosts only for immature tick stages would cause an increase of tick population, resulting in an overall amplifying effect on the transmission of tick borne agents \citep{Randolph2012,Mannelli}. Accordingly, the experimental removal of lizards in an area in the Western USA, where these reptiles are important hosts for immature \textit{I. pacificus}, but are non-competent as reservoirs for \textit{B. burgdorferi}, unexpectedly resulted in a decrease of the density of infected nymphal ticks \citep{Swei2011}. Furthermore, \textit{B. afzelii} was relatively frequent in a study area in Germany where lizards were identified as reservoirs for \textit{B. lusitaniae} \citep{richtermatuschka2006}.
		
		Field observations as well as mathematical models suggest that it is only above certain population density thresholds that non competent hosts cause dilution of tick borne agents \citep{Rosà2003,Rosà2007,Cagnacci2012,Bolzoni2012}. Consequently, the role of lizards on the transmission dynamics of \textit{B. afzelii} on Le Cerbaie and, more in general, in the Mediterranean area, should be further investigated. This might be accomplished either by identifying locations with naturally different lizard densities, or through experimental manipulation of lizard populations.  Furthermore, since amplification of transmission of \textit{B. burgdorferi} sl by vector augmentation by lizards was not considered in our model, such an effect should also be investigated from the theoretical point of view. 
		
	\section*{Acknowledgements}
			LF acknowledges support from the Lagrange Project, CRT and ISI Foundation. MG acknowledges funding from local research grants of the Universit\`a degli Studi di Torino.
		
	\appendix
	\section{$\hat{x}$ as function of $h_R$}\label{function}
		We have two, coincident, ways to explicitly solve Eq. \ref{eq_eq_a} (and the corresponding equation for \textit{B. lusitaniae}) for $h_R$ ($h_L$):
			\begin{itemize}
				\item numerically solve them
				\item Eq. \ref{eq_eq_a} could be explicitly solved for $h_R$. Then 
				\begin{equation}
					h_R=\Phi(\hat{x})=-\frac{1}{\hat{x}} \log\left(1-\frac{\hat{x}-\left(1-\mu_R\right)^2\cdot \hat{x}}{1-\left(1-\mu_R\right)\cdot \hat{x}}\right)
				\end{equation}
				$\Phi(x)$ for $x\neq 1$ is continuous bijective function. Therefore, for $h_R\in[0,1]$, is possible to invert it and to have $\hat{x}=\Phi^{-1}(h_R)$. In a similar fashion we could have $\hat{y}=\Gamma^{-1}(h_L)$.
			\end{itemize}

\bibliography{biblio}

\end{document}